\documentclass[twocolumn,nofootinbib,superscriptaddress]{revtex4-1}

\usepackage[english]{babel}
\usepackage[OT1]{fontenc}
\usepackage[latin1]{inputenc}
\usepackage{xcolor}
\usepackage{latexsym,amsmath,amsfonts,amssymb,mathtools,mathrsfs}
\usepackage{footmisc} 
\usepackage{hyperref}

\newcommand\dimM{D}
\newcommand\mP{M_{\mathrm{P}}}
\newcommand\CosmC{\Lambda_0}

\definecolor{oucrimsonred}{rgb}{0.6, 0.0, 0.0}

\begin{document}
\title{Comment on ``Einstein-Gauss-Bonnet  Gravity in  Four-Dimensional Spacetime"}

\author{Julio Arrechea}
\email{arrechea@iaa.es}
\affiliation{Instituto de Astrof\'isica de Andaluc\'ia (IAA-CSIC), Glorieta de la Astronom\'ia, Granada, Spain}

\author{Adri\`a Delhom}
\email{adria.delhom@uv.es}
\affiliation{Departamento de F\'{i}sica Te\'{o}rica and IFIC, Centro Mixto Universidad de Valencia - CSIC. Universidad de Valencia, Burjassot-46100, Valencia, Spain}

\author{Alejandro Jim{\'e}nez-Cano}
\email{alejandrojc@ugr.es}
\affiliation{Departamento de F\'{i}sica Te\'{o}rica y del Cosmos and CAFPE, Universidad de Granada, 18071 Granada, Spain}

\date{\today}

\maketitle

In this Comment, we elaborate on several points raised in Ref. \cite{Glavan2020}. The authors claimed to have found a four-dimensional gravitational theory which fulfills the assumptions of the Lovelock theorem \cite{Lovelock1972} though not its implications. To that end, they employed a \textit{regularization} procedure already outlined in Ref. \cite{Tomozawa2011}. This procedure consists of rescaling the coupling constant of the Gauss-Bonnet (GB) term by $1/(D-4)$ and taking the $D\to4$ \textit{limit} after varying the Einstein-Gauss-Bonnet (EGB) action. The authors of Ref. \cite{Glavan2020} claim that the variation of the GB term is proportional to $(D-4)$, canceling the $1/(D-4)$ factor and hence yielding a nonvanishing contribution to the field equations in $D=4$. This claim does not stand a thorough analysis given that the variation of the $k$th order Lovelock Lagrangian can be decomposed as \cite{Arrechea2020,Gurses2020}
\begin{equation}
\frac{1}{\sqrt{|g|}}\frac{\delta S^{(k)}}{\delta g^{\mu\nu}} = (\dimM-2k) A^{(k)}_{\mu\nu} + W^{(k)}_{\mu\nu}, \label{eq: AW decomposition}
\end{equation}
where $W^{(k)}_{\mu\nu}$ is a tensor from which no $(D-2k)$ can be extracted and which does not vanish for a general $D>2k$. The GB term is given by $k=2$, and the field equations proposed in Ref. \cite{Glavan2020} for arbitrary dimension are
\begin{equation}\label{fieldeqs}
G_{\mu\nu}+\frac{1}{\mP^2}\CosmC g_{\mu\nu}+\frac{2\alpha}{\mP^2}\left(A^{(2)}_{\mu\nu}+\frac{W^{(2)}_{\mu\nu}}{\dimM-4}\right)=0\,,
\end{equation}
where the explicit form of $A^{(2)}_{\mu\nu}$ and $W^{(2)}_{\mu\nu}$ can be found in Ref. \cite{Arrechea2020}.
Although the $1/(D-4)$ rescaling compensates the factor multiplying the $A^{(2)}_{\mu\nu}$ term, the same procedure renders the $W^{(2)}_{\mu\nu}$ term indeterminate. This owes to the fact that, although $W^{(2)}_{\mu\nu}$ vanishes in $D=4$, it does so due to algebraic reasons \cite{Gurses2020,Arrechea2020} and not because it is proportional to some power of $(D-4)$. Hence, the indeterminate term $W^{(2)}_{\mu\nu}/(D-4)$ renders the field equations \eqref{fieldeqs} ill-defined in four dimensions. Indeed, the first problem to address would be to make sense of the limit of a tensor field \cite{Gurses2020,Arrechea2020,Mahapatra2020,Ai2020,ZanelliPriv}, since these objects are defined for integer values of $D$ only.

Despite the above discussion, the field equations \eqref{fieldeqs} are well defined in $D=4$ when constrained to particular geometries in which $W^{(2)}_{\mu\nu}$ vanishes in arbitrary dimensions. This is the case, for instance, for all conformally flat geometries, including the Friedmann-Lemaitre-Robertson-Walker or maximally symmetric solutions found in Ref. \cite{Glavan2020}. However, metric perturbations will be sensible to the ill-defined terms, hence rendering these solutions unphysical. Indeed, though linear perturbations around a maximally symmetric background are oblivious to these pathologies \cite{Glavan2020}, they enter at second order through terms proportional to $1/(D-4)$ \cite{Arrechea2020}. In this direction, other works also showed that an infinitely strongly coupled new scalar degree of freedom appears beyond linear order \cite{Bonifacio2020}. These results strongly suggest that Eq. \eqref{fieldeqs} is generally ill-defined in four dimensions.

A possible way to circumvent the pathologies of the above field equations \eqref{fieldeqs}  would be to get rid of the $W^{(2)}_{\mu\nu}$ term.  However, $A^{(2)}_{\mu\nu}$ is not divergenceless \cite{Arrechea2020}. Hence, by virtue of the Bianchi identity under diffeomorphisms, we conclude that there is no diffeomorphism-invariant action whose variation gives Eq. \eqref{fieldeqs} without the $W^{(2)}_{\mu\nu}$ term.

 Regarding spherically symmetric metrics of the form
\begin{equation}\label{sphericallysymmetricmetric}
    {\rm d}s^2 = B(r){\rm d}t^2 -B^{-1}(r){\rm d}r^2-r^2d\Omega^2_{D-2}\,,
\end{equation}
	for them to have a vanishing $W^{(2)}_{\mu\nu}$ in arbitrary dimension, there are conditions that $B(r)$ must fulfill \cite{Arrechea2020}. The spherically symmetric geometries presented in Ref. \cite{Glavan2020} do not satisfy these requirements and, therefore, cannot be solutions of Eq. \eqref{fieldeqs}, since this is not a well-defined set of equations in this case.  Indeed, as proven in Ref. \cite{Arrechea2020}, they are neither solutions of the truncated equations \eqref{fieldeqs} without the $W^{(2)}_{\mu\nu}$ term. This is not surprising, since the authors of Ref. \cite{Glavan2020} derived these geometries by taking $D=4$ in the $D\geq5$ solutions for the EGB theory found in Ref. \cite{Boulware1985} and then rescaling the GB coupling by $1/(D-4)$, instead of finding a solution to the rescaled field equations \eqref{fieldeqs} in $D=4$. Furthermore, the authors of Ref. \cite{Glavan2020} state that the central curvature singularity of these geometries cannot be reached by an observer. Nevertheless, as shown in Ref. \cite{Arrechea2020}, radial freely falling observers do reach the singularity in finite proper time, contradicting this claim.

\end{document}